\documentclass[letter]{ptptex}

\title{
Decoherence of a Spin Qubit Coupled with a Spin Chain
}

\author{
Toshifumi \textsc{Itakura} \footnote{E-mail: itakurat@s6.dion.ne.jp}
}

\inst{
 Department of Physics, 
 Kyoto University,
 Kyoto 606-8502, Japan \\
}

\recdate{
October 25, 2004}

\abst{
In this report, we examine the relaxation phenomena of a spin 
   qubit coupled to a spin chain with a $1/r^2$ interaction.
}

\begin{document}
\maketitle

\section{Introduction}

Among various proposals for quantum computations,
 quantum bits (qubits) in solid state materials,
 such as superconducting Josephson junctions
\cite{Nakamura},
           and quantum dots
\cite{Hayashi,Tanamoto,Loss},
           have the advantage of  scalability.
Such coherent two level systems constitute qubits and
 a quantum computation can be carried out as a
        unitary operation appied to multiple qubit systems.
It is essential that this quantum coherencre be 
       maintained during computation.
However, dephasing is difficult to 
     avoid, due to the system's interaction with
     the environment.
The decay of off-diagonal elements of
         the qubit density matrix signals  the occurrence of dephasing.
Various environments can cause dephasing.
In solid state systems, the effect of phonons is ubiquitous
\cite{Fujisawa_SC}.
The effect of electromagnetic fluctuation
     has been extensively studied for Josephson junction charge qubits
\cite{Schon}.

The fluctuations of the nuclear spins of impurities can also
 be a cause of dephasing.
It has recently been shown experimentally that the coupling 
      between the spin of an electron in a quantum dot 
      and the environment is very weak
                  \cite{Fujisawa_Nature,Fujisawa_PRB,Erlingsson,Khaetskii}.
For this reason, the dephasing time 
      of a spin qubit conjectured to be very long.
However, both donor impurities
       and nuclear spins in semiconductors
       \cite{Ladd}
       have been suggested as possible building blocks for feasible
       quantum dots architectures.
The proposal based on experimental findings of quantum computation  using 
 an Si$^{29}$ array
 is another possibility
   \cite{Ladd}.
For a spin qubit system in 2D, the coupling between nearest neighbor
 spin chain can cause a dephasing.
When an interaction is between the qubits themselves,
   it can in principle be incorporated into 
   the quantum computer Hamiltonian,
   although this would lead to more complicated gate sequences.
Therefore it is instructive to analyze the error 
   introduced by ignoring some
   of these interactions,
   as done in the case of dipolar coupled spin qubits
   \cite{Sousa_1}.
For a quantum spin chain with a $1/r^2$ interaction,
an exact expression of the dynamical correlation function has been obtained
\cite{Haldane}.
Using this expression,
we consider the case in which  one qubit is coupled to each spin of 
the spin chain. 
We examine the relaxation phenomena of a spin qubit
in array  that is coupled to a spin chain 
     with long-range interactions.
Integrating over the spin chain variables,
we obtain the influence function of the qubit system.
Using this influence function,
we examine the dephasing of the qubit's density matrix.
In the present study, 
   we especially concentrated on 
   the correlation effect between  
   the spin of the qubit and that of the spin chain,
   and show how the characteristic nature of environment appears.
The result is that the dephsing rate increases as a function of the 
intrachain interactions,
because the fluctuations of the spinon is suppressesd interaction increases.
\section{Hamiltonian}
   
We examine the Hamiltonian which is given by,
\begin{eqnarray}
   \label{eqn:H1}
{\cal H} &=& H_{qb}  + H_{qb-{\rm spin}} + H_{{\rm spin}} \\
   \label{eqn:H2}
  H_{qb} &=& \hbar \Delta I_z, \\  
  H_{{\rm int}} &=& 
  \gamma_N \hbar^2 ( A_{zz} I_z S_{z0} 
  + A_{\perp} ( I_+ S_{-0} + I_{-} S_{+0})),
   \\
   \label{eqn:H4}
  H_{{\rm spin}} &=& J \sum_{i,j=N}^N [d(x_i - x_j)]^2
  {\bf S}_i \cdot  {\bf S}_j. 
\end{eqnarray}
  where N is the number of sites, and 
$d(n) = (N/\pi) \sin ( \pi n /N) \rightarrow n$
       as $N \rightarrow \infty$.
        Also  $i$ is the index of  $N$ lattice.
$\Delta$ is the effective magnetic field applied to a  spin qubit,
        and 
         $ A_{zz} \gamma_N \hbar$
         is  the magnitude of the fluctuations 
         in the coupling constant.
A system of this kind has also been examined in the context of
          a giant spin that is coupled to a surrounding nuclear spin
          interacting bilinearly with a spin bath in a 
          magnetic field and with internal interactions of
          canonical form
\cite{QTM,TCS}.
In this study we examine the dephasing behavior
  by using an exact dynamical correlation function
  and examine the characteristic nature of a bath
  that interacts with itself.

\section{Influence function}
We examine first a pure dephasing event ($A_{\perp} = 0$),
and then we study  dephasing with a dissipation event
 ($A_{\perp} \ne 0 $).
The density matrix of spin qubits is given by
\begin{equation}
  \rho ( I_{z+}^f, I_{z-}^f )
   = \int^{I_{z+}(t) =I_{z+}^f, I_{z-} (t) =I_{z-}^f}_{ I_{z+}(0)=I_{z+}^i, I_{z-} (0) = I_{z-}^i} 
  [ d I_{z+} ] [ d I_{z-} ] 
  {\exp} ( \frac{i}{\hbar} (I_{qb} [ I_{z-} ] - I_{qb} [ I_{z+} ] ))
  F[I_{z+},I_{z-}],
\end{equation}
where $I_{qb} [I_{\pm}] = \int_0^t \hbar \Delta I_{z \pm}$, and 
the influence function (IF) is defined as
\begin{equation}
 F[ I_{z+}, I_{z-} ] = \int [ d   {\bf S}_{+i}  ] [ d  {\bf S}_{-i} ]
 \delta (  {\bf S}_{+i} (t) - {\bf S}_{-i}(t) ) 
  \rho (  {\bf S}_i  (0),  {\bf S}_j (0) ) 
 {\rm exp} \{ 
 \frac{i}{\hbar} (  I [ S_+ ]- I [ S_- ] ) 
\} .
\end{equation}
For the following investigation, we assume $A_{\perp}$=0 and 
that the environment system action is given by
 $ I[ {\bf S} ] = I_0 [ {\bf S} ] + I_{int} [ I_{z}, S_z^0 ]$,
 with free part
\begin{eqnarray}
   I_0 [  S ] &=& \frac{ (A_{zz} \gamma_N \hbar^2)^2}{4 \hbar^2} 
   \int_0^t  \int_0^{t_1} dt_1 dt_2  \sum_{i.j}  {\bf S}_i (t_1)
    {\bf \Delta}_{00p}^{-1} ( t_1 , i , t_2 , j) 
    {\bf S}_j ( t_2 ) \nonumber \\
   &+& i \frac{\theta}{4\pi} \int dx \int_0^{t} dt {\bf S} 
        \cdot (\frac{\partial {\bf S}}{\partial x} 
       \times \frac{\partial {\bf S}}{\partial t} ) 
\end{eqnarray}
where $\theta=2 \pi n $ and ${\bf \Delta}_{00p} (t_1,i,t_2, j) $
 is the free propagator of the environmental system
  at zero temperature, which is defined on a closed time path and has
  four components:
 \cite{Chou}
\begin{eqnarray}
  \label{eqn:Green}
  {\bf \Delta}_{00p}  (i, t_1, j, t_2) 
  = \left(
   \begin{array}{cc}
  {\bf \Delta}_{00}^{++} (i, t_1,j, t_2) & {\bf \Delta}_{00}^{+-}  
  (i, t_1, j, t_2) \\
  {\bf \Delta}_{00}^{-+}  (i, t_1,j, t_2) & {\bf  \Delta}_{00}^{--} 
  (i, t_1, j, t_2) 
  \end{array}
  \right). \nonumber \\
\end{eqnarray}
When $A_{\perp}=0$,
${\bf \Delta}_{00p} \rightarrow \Delta_{00p}^z$,
only the $z$ component of the qubit is important.
For the Berry phase term, because
  n is odd, the environment is a one-dimensional spin-1/2 system
  \cite{Zagoskin}.
  In the present model, the spin chain is a gapless solvable model 
  (a special point in spin chain systems).
  Therefore, we can completely  integrate over the spin chain degree of freedom
  and the Berry phase term does not make system to be distempered.
Thus, introducing the incoming interaction picture for the environment,
   and integrating out the spin chain environment,
   we can easily verify that Eq. (\ref{eqn:Green}) becomes 
   \cite{Chou},
\begin{eqnarray}
  F[ I_{z+}, I_{z-}] &=& 
   {\rm exp} [ - i \frac{(A_{zz} \gamma_N \hbar^2)^2}{4 \hbar^2} 
    \int_{0}^{t} \int_{0}^{t_1} dt_1 dt_2 \nonumber \\  
   &&  ( I_{z+} (t_1) \Delta_{00}^{z++} (0,t_1,0,t_2) I_{z+} (t_2) 
    + I_{z-} (t_1) \Delta_{00}^{z--} (0,t_1,0,t_2)
     I_{z-} (t_2) \nonumber \\ 
    &-& I_{z+} (t_1) \Delta_{00}^{z+-} (0,t_1,0,t_2) I_{z-} (t_2) 
    - I_{z-} (t_1) \Delta_{00}^{z-+} (0,t_1,0,t_2) I_{z+} (t_2) )].\nonumber \\    
\end{eqnarray}
For convenience we change  the coordinates,
$  \eta \equiv (I_{z+} + I_{z-})/2, 
  \xi \equiv (I_{z+} - I_{z-})/2. $
The coordinates of $\eta$ and $\xi$, given by $\eta$ and $\xi$ are called the "sojourn" and "blip".
In terms of these variables, the density matrix is described by 
$ \rho (\eta =1) = |\uparrow \rangle  \langle \uparrow|,
 \rho ( \eta =-1) = |\downarrow \rangle \langle \downarrow|, 
 \rho (\xi  = 1 ) =  |\uparrow \rangle \langle \downarrow|,
 \rho ( \xi = - 1)  = |\downarrow \rangle \langle \uparrow |.$

Then, the IF becomes
\begin{eqnarray}
 F[ \eta, \xi ] &=&
  {\rm exp} [ - i \frac{( A_{zz} \gamma_N \hbar)^2}{4}
 \int_{0}^{t} \int_{0}^{t_1} dt_1 dt_2  
 \{ \xi (t_1)  G^R (t_1,0,t_2,0) \eta (t_2) \nonumber \\ 
                      &+& \eta (t_1) G^A (t_1,0,t_2,0) \xi (t_2)  
                      - \xi (t_1) G^K (t_1,0,t_2,0) 
                      \xi (t_2) \}  ], 
\end{eqnarray}
where $G^R (t_1,0,t_2,0)$, $G^A (t_1,0,t_2,0)$ 
 and the $G^K (t_1,0,t_2,0)$ are
 the retarded Green function, advanced Green function 
 and Keldysh Green function for $i=j=0$.
 It should be noted that the starting Hamiltonians are those given
  Eqs.(\ref{eqn:H1}) $\sim$ (\ref{eqn:H4})
  \cite{Chou}. 
Because the propagator of a qubit has no time dependence,
the blip state and sojourn state do not change in time.
Therefore, when we choose the initial condition of the qubit density matrix 
  to be a coherent state,
  such as 
$   \rho ( t=0) = 
                \pm ( |\uparrow \rangle \langle \downarrow| \pm |\downarrow \rangle \langle \uparrow|) ,$
 time evolution occurs only in the off-diagonal channel.
 In addition, even if we start from an off-diagonal state,
the interaction Hamiltonian does not allow a spin flip process.
Hence the blip state does not change in time.
Therefore we can set $\xi(t)=\xi (=\pm 1)$ and $\eta(t)=0$ for all $t$.
This leads to the following exact expression of the dephasing rate :
\begin{eqnarray}
 \rho (\xi,t) &=&
  {\rm exp} [ i \Delta t - 
 i \frac{( A_{zz} \gamma_N \hbar )^2}{4}
 \int_{0}^{t} \int_{0}^{t_1} dt_1 dt_2   G^K (t_1,0,t_2,0) ] \nonumber \\  
  &=& {\rm exp} [ i \Delta t -  \frac{(A_{zz} \gamma_N \hbar)^2}{4}
 \int_{0}^{t} \int_{0}^{t_1}
 dt_1 dt_2  \{ S^z_0 (t_1), S^z_0 (t_2) \} ] .  
\end{eqnarray}
Here, $ \{ S^z_0,S^z_0 \} $ is the symmetrized correlation function,
  defined by $\{ A,B \} =AB + BA$, because the Keldysh Green function is
  a symmetrized correlation function  for 
  the base of the present Hilbert space
  \cite{Zagoskin}.

\section{Results}
First, we examine a pure dephasing event.
(i.e., $A_{\perp}$=0)
Then using analytical expressions, the time evolution of 
the off-diagonal density
   matrix is given by
\begin{eqnarray}
   \label{eqn:rho}
  - \ln \rho(\xi = \pm 1,t )
   &=& i \Delta t + \frac{( A_{zz} 
  \gamma_N \hbar^2)^2}{4 \hbar^2}
  \int_{0}^{t} \int_0^{t_1} dt_1 dt_2  \{ S^z_0 (t_1), S^z_0 (t_2) \}. 
  \nonumber 
\end{eqnarray}
Here, $E=E(\lambda_1.\lambda_2)=\frac{\pi v}{2}
 (\lambda_1^2 + \lambda_2^2 -2
  \lambda_1^2 \lambda_2^2)$, where $v$ is the spinon velocity.
We carried out the numerical estimation at zero temperature.

In the initial regime, $\frac{\pi v t}{2} \ll 1 $, we have
$  - {\rm Re} \ln \rho(t)
   = \frac{(A_{zz} \gamma_N   \hbar)^2}{8} t^2. $
This is Gaussian behavior which is displayed in the initial regime.
This comes the memory effect of the spin chain reservoir.
In the stationary regime, we have
$  - {\rm Re} \ln \rho (t)
   = \frac{( A_{zz} \gamma_N  \hbar)^2}{8v} t.$
This behavior leads to exponential decay of the off-diagonal element of 
the density matrix in the long time tail.
Therefore, after tracing out the spin chain system, the  qubit density matrix
exhibits a maximal mixed state, i.e., an  entangled state is formed.
In the intermediate regime ($1<\frac{\pi v t}{2} < 50$)
, the dephasing rate exibits  
oscillatory behavior.
Due to the restriction of the spin magnitude of the spin chain,
 during the transient regime between decoupled state and an entangled state
  there exists oscillatory behavior.
In the stationary limit ($ 50 \ll \frac{\pi v t}{2} $),
 the maximally entangled state is formed.
Therefore, after we trace out the spin chain system, shows
the qubit density matrix is in the maximally mixed state.
The time traces qubit and an spin chain are separable state  to 
   maximally entangled state .

The disappearance of the diagonal elements of
 the qubit indicate this behavior.   
In another context this behavior represents an effect of 
the qubit system on the
environment.

We now examine the finite temperature behavior.
In the frequency regime,
we carry out the Fourier transform of the integral in Eq. (\ref{eqn:rho}).
Then, the expression for the of dephasing rate is 
\begin{equation}
  -{\rm Re} \ln \rho (t) = \frac{(A_{zz} \gamma_N \hbar)^2}{4} 
  \int_{-\infty}^{\infty}
  d \omega 
   \rangle \{ S_z (0,\omega),S_z (0,-\omega) \} \langle 
   \left( \frac{\sin( \omega t / 2 )}{\omega/2} \right)^2, 
\end{equation}
where $T=0$.

At finite temperature, because generally a spin-1/2 system obeys Fermi statistics
we can obtain the dephasing rate by multipling the 
  to spectral weight function by $\tanh(\omega/2 k_B T)$,
\begin{equation}
    \label{eqn:HSS}	
  - {\rm Re} \ln \rho (t) = \frac{(A_{zz} \gamma_N \hbar)^2}{4} 
  \int_{-\infty}^{\infty}
  d \omega \tanh (\frac{\omega}{2 k_B T})
  \langle \{S_z (0,\omega),S_z (0,-\omega)\} \rangle \left(
   \frac{\sin( \omega t /2 )}{\omega/2} \right)^2 .
\end{equation}
The validity of the expression is demonstrated in by Ref. \citen{Chou},
and it should be required because
 of antiperiodic conditions (Fermi statistics).
This expression is reminiscent of  the Ingold-Nazarov theory
in quantum tunneling events.
\cite{Leggett,Odintsov,Nazarov}.
They showed the relation $P(t) = {\rm exp} K(t)$, with
$ K(t) = \frac{4}{\pi \hbar} \int_0^{\infty} \frac{J(\omega)}{\omega^2}
     ( \coth \left( \frac{\hbar \omega}{2 k_B T} \right)
      [ ( \cos( \omega t) -1))
     - i \sin \omega t ]) ,$
where $J(\omega)$ is the bath spectral density and $P$ is a coherence function,
which represents the Coulomb-blocked effect.
From this correspondence,
 we can conclude that the bath spectral density function of
 the Haldane-Shastry model at finite temperature, is given by
\begin{equation}  
 J(\omega)_{HS} \equiv
  \frac{\pi \hbar}{8} \tanh(\omega/2 k_B T)^2 
 (A_{zz} \gamma_N \hbar)^2 
 \langle \{ S_z (0,\omega), S_z (0, -\omega) \} \rangle .
\end{equation} 
At finite temperatures, the spin bath has a smaller effect 
  on the system, because
   of the possibility for saturation of the populations in the bath.
A conspicuous counterexample to a spin bath is the case in which the ladder of
   excited states of a nonlinear bath mode  narrows with increasing
   energy.
In this case, we find enhancement of $J(\omega)$
 in comparison with the spectral density of a harmonic bath.    
This type of temperature dependence
   results from the statistics of the bath  
\cite{Weiss,QTM,TCS}.
At high temperature spins of electrons behave
  like those of classical particles.
However, in spite of this fact, such classical behavior had not been studied to this time.

Next, we examine the situation in which   
  dephasing with dissipation occurs ($A_{\perp} \ne 0$).
In the present case, the second-order perturbation approximation 
is not accurate,
although for a pure dephasing event we obtain an exact result.
In  previous studies, Fermi's Golden Rule was  used.
In this case, the transverse relaxation, $T_1^{-1}$  and 
  the   longitudinal transition rate are given by
     \cite{Sachdev}
$  - \ln ( \eta = 1,t) = T_1^{-1}
    = \frac{(A_{zz} \gamma_N \hbar)^2}{4 v}
   ( 3 +\frac{9}{2 \pi^2} \left( \frac{\Delta}{v} \right)^2 )$ 
 and $ - \ln ( \xi = 1 ,t)
    = \frac{1}{2} T_1^{-1} 
  + \frac{1}{8 v} ( A_{\parallel} \gamma_N \hbar)^2.$
The results show that the dephasing suppressed
 as the strength of the intrachain interaction increases.
This is because the fluctuations 
   of a spinon are suppressed as $J$ increases.
Also, as $\Delta$ increases,
  the dephasing increases,
  because the spin bath behaves like as ohmic bath.
It is conjectured that for a spin bath, the higher-order 
 corrections of $\Delta$
 are negative.
This behavior leads us to hypothesize a finite band width of $\frac{\pi v}{2}$.
In the present model,
   we can confirm our results, because
  the universality class of the Haldane-Shatry model
   is  the Tomonaga-Luttinger liquid with $K=1/2$
   \cite{Kawakami},
   and  at zero temperature, the dephasing rate of dephasing
    with dissipation is constant
    \cite{Kawaguchi}.

\section{Summary}
In summary, we had examined the dephasing rate of a spin qubit system coupled
   with a spin chain.
Due to the fluctuations of the spin of the spin chain, 
the dephasing is supperessed as $J$.
The time dependence exhibits oscillatory behavior, which comes from
   the entangled state.
We also examined  dephasing with dissipation.
We found that the dephasing with dissipation is suppressed as
 the strength of the magnetic field
   applied to qubit system (energy width of two-level system) is decreased.
It should be noted that for pure dephasing, our results are exact and
  can be applied to other bath systems by using the
  bath spectrum density function.

{\bf Acknowledgements} 
The author thanks Akira Kawaguchi, Yasuhiro Tokura, 
Naoto Ogata, Akira Furusaki, 
   Norio Kawakami, R. Fazio and H. Kawai 
for their advice and stimulating discussions.
This work is supported by a Grant-in-Aid for the 21st Century COE "Center for Diversity and Universality in Physics" from the Ministry of Education, Culture, Sports, Science and Technology (MEXT) of Japan.

\end{document}